\documentclass[%
 aip,
 amsmath,amssymb,
 reprint,%
]{revtex4-1}

\usepackage{graphicx}
\usepackage{dcolumn}
\usepackage{bm}

\usepackage[utf8]{inputenc}
\usepackage[T1]{fontenc}
\usepackage{mathptmx}
\usepackage{etoolbox}

\usepackage{color}

\makeatletter
\def\@email#1#2{%
 \endgroup
 \patchcmd{\titleblock@produce}
  {\frontmatter@RRAPformat}
  {\frontmatter@RRAPformat{\produce@RRAP{*#1\href{mailto:#2}{#2}}}\frontmatter@RRAPformat}
  {}{}
}%
\makeatother
\begin{document}

\preprint{AIP/123-QED}

\title{Tuning cryogenic Jahn-Teller transition temperatures in magnetoelectric rare earth vanadates}
\author{Kejian Qu}
 \affiliation{Department of Physics and Materials Research Laboratory,\\ University of Illinois at Urbana-Champaign, Urbana, IL 61801, USA}
\author{Daniel P. Shoemaker}%
 \email{dpshoema@illinois.edu}
\affiliation{ 
Department of Materials Science and Enginerring and Materials Research Laboratory, University of Illinois at Urbana-Champaign, Urbana, IL 61801, USA}

\date{\today}

\begin{abstract}
Few materials undergo cooperative Jahn-Teller (JT) transitions at low temperatures, but zircon-type oxides are one class that includes DyVO$_4$, which transforms from a tetragonal to an orthorhombic structure at around 13.6~K, with a narrow transition temperature range within 0.5~K.
Since many rare-earth ions can be accommodated in the structure, there should be ample routes to vary the transition temperature and structural effects of the transition. 
We have synthesized pure DyVO$_4$ and solid solutions Dy$_{1-x}$Tm$_x$VO$_4$ ($x$ = 0.05, 0.1, 0.15, 0.2, and 0.5) and Dy$_{1-y}$Pr$_y$VO$_4$ ($y$ = 0.03, 0.05, 0.1, 0.2, and 0.5), all by solution precipitation.
X-ray diffraction shows a systematic peak shift and a linear change of lattice parameters with increasing substitution.
We demonstrate through heat capacity measurements that both Tm$^{3+}$ and Pr$^{3+}$ substitutions cause a depression of the cooperative JT transition temperature.
When the substitution level increases, the JT transition is eventually suppressed. 
We examine a mean-field approximation model that can explain the both the JT transition temperature shift and its eventual disappearance. 
The cooperative JT effect in these zircon-type oxides is known to exhibit a large magnetoelectric response, which should follow from the mean-field behavior. 
The structural transformation can be easily detected via diffraction and can be used as a temperature calibration in low temperature experiments.
\end{abstract}

\maketitle

The Jahn-Teller (JT) effect\cite{JT-review} distorts the crystal lattice and breaks the symmetry in solid-state materials spontaneously: degenerate or nearly degenerate electronic states will distort to create an energy gap below a critical temperature.
The splitting of degenerate electronic levels can lower the total electronic energy for ions with partially filled shells, such as Mn$^{3+}$ in octahedral sites\cite{Mn-JT} and Fe$^{2+}$ in tetrahedral sites.\cite{Fe-JT} 
This energy gain outweighs the elastic strain energy penalty of the lattice distortion.
The cooperative JT effect\cite{Gehring-long, Gehring-short} occurs in crystalline systems where a local distortion in each complex with a JT-active ion couples cooperatively through site-to-site interaction, so that the whole crystal or grain will distort along a specific crystallographic direction.

As classical examples exhibiting cooperative JT transition, rare earth vanadates (REVO$_4$) have been synthesized\cite{German} and studied by X-ray diffraction (XRD)\cite{whole-group-old} and neutron diffraction.\cite{whole-group-new-TmVO4-PrVO4}
Rare earth vanadates share the tetragonal zircon-type structure.\cite{rare-earth-vanadates-structure}
Among them, TmVO$_4$,\cite{TmVO4-1, TmVO4-2} TbVO$_4$,\cite{TbVO4-1, TbVO4-2} and DyVO$_4$\cite{DyVO4-JT-Cooke, DyVO4-longlist} can undergo cooperative JT transition and become orthorhombic at around 2~K, 34~K, and 14~K, respectively.
Other isostructural zircon-type compounds exhibit cooperative JT transitions, as long as there are degenerate electron states, such as DyAsO$_4$,\cite{DyAsVO4} TmAsO$_4$,\cite{TmAsVO4} TbAsO$_4$,\cite{TbAsO4} TbPO$_4$,\cite{TbPO4} and DyCrO$_4$\cite{DyCrO4-structure, DyCrO4-temperature-magnetocaloric} at around 11~K, 6~K, 28~K, 2~K, and 31~K, respectively.

\begin{figure}
    \centering
    \includegraphics[width=\columnwidth]{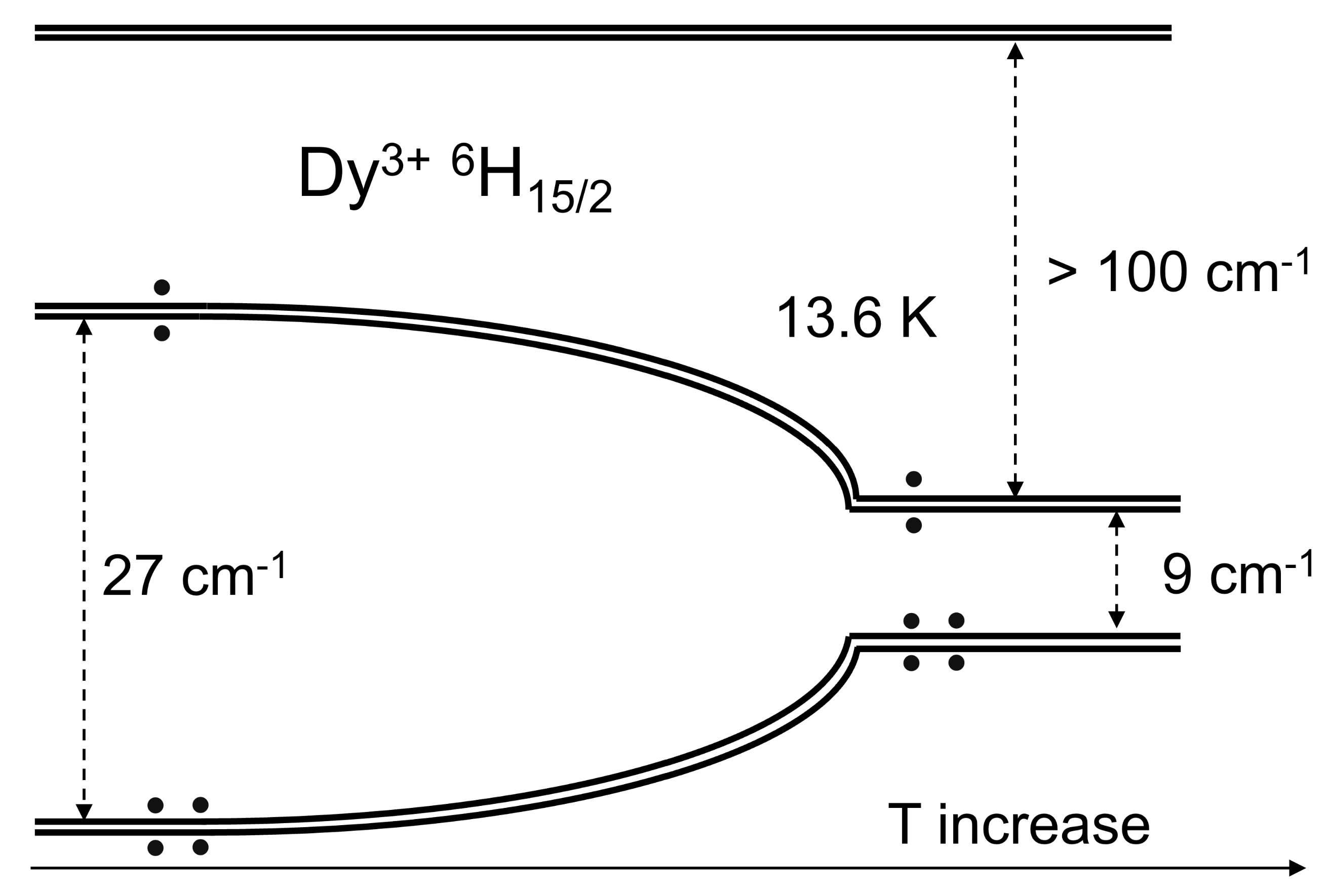}
    \caption{Lower 4f electronic states of Dy$^{3+}$ in its ground state $^6$H$_{15/2}$ for DyVO$_4$. Electrons are denoted by dots, and their total energy will be lower when JT transition happens.}
    \label{fig: DyVO4-levels}
\end{figure}

Among these rare-earth vanadates, DyVO$_4$, which has been widely studied,\cite{DyVO4-longlist} undergoes a cooperative JT transition from a tetragonal zircon structure $I$4$_1$/$amd$, or $\mathrm{D^{19}_{4h}}$ with lattice parameters $a=7.1395$~{\AA} and $c=6.2899$~{\AA}, to an orthorhombic structure $Imma$, or $\mathrm{D^{28}_{2h}}$ with lattice parameters $a=7.1515$~{\AA}, $b=7.1275$~{\AA} and $c=6.2899$~{\AA}.\cite{DyVO4-lattice-parameters-splitting, DyVO4-JT-harmonic-oscillator-theory}
In the ground state configuration $^6$H$_{15/2}$ of the Kramers ion Dy$^{3+}$, the lowest two 4f Kramers doublets (4 states in total) are involved in the cooperative JT transition. They have a spacing of only 9~cm$^{-1}$ (1.1~meV), almost degenerate, while the next state is more than 100~cm$^{-1}$ above and it will not participate in the JT transition.\cite{Raman-review}
A diagram of relevant electronic states of Dy$^{3+}$ is shown in Figure \ref{fig: DyVO4-levels}.\cite{DyVO4-levels-1, DyVO4-levels-2}
Heat capacity measurements are widely seen to show an upward anomaly when the material is cooled across its JT transition temperature, due to the availability of new energy levels after splitting. Previous studies\cite{DyVO4-heat-capacity-Ising, DyVO4-heat-capacity-2, DyVO4-heat-capacity-3} have shown the JT transition temperature of DyVO$_4$ to be around 14~K when no magnetic field is present.

These JT-active zircon-type vanadates are also well-known for their magnetoelectric effects.\cite{Magnetoelectricity-theory} For DyCrO$_4$ it has been shown that the transition temperature can be shifted by doping with Bi$^{3+}$ for Dy$^{3+}$, together with the suppression of magnetoelectric effects.\cite{DyCrO4-magnetoelectric} For DyVO$_4$ it has been shown that a varying magnetic field can shift the transition temperature.\cite{DyVO4-magnetoelectric}
Here we seek to explain how to predict the magnitude of the magnetolectric response (related to the heat capacity anomaly) and the transition temperature (necessary for magnetocaloric applications, or to use these compounds as calibration standards for low-temperature XRD measurements). 

We demonstrate through XRD and Rietveld refinement that pure DyVO$_4$ and solid solutions Dy$_{1-x}$Tm$_x$VO$_4$ ($x$ = 0.05, 0.1, 0.15, 0.2, and 0.5) and Dy$_{1-y}$Pr$_y$VO$_4$ ($y$ = 0.03, 0.05, 0.1, 0.2, and 0.5) were successfully synthesized by solution precipitation.
We also demonstrate the tuning of cooperative JT transition temperatures from around 14~K down to around 8~K as the Tm$^{3+}$ or Pr$^{3+}$ substitution level increases and the complete suppression of JT transition when the substitution level is high enough.
It is uncommon for such cooperative structural phase transformations to occur at cryogenic temperatures, and it is even more rare that the transition temperatures can be tuned with precision.

Pure DyVO$_4$ and solid solutions Dy$_{1-x}$Tm$_x$VO$_4$ and Dy$_{1-y}$Pr$_y$VO$_4$ were all synthesized by solution precipitation, with detailed synthesis procedures in Supplementary Materials.\cite{supplement}
The advantages of solution precipitation synthesis include uniform mixing of rare earth cations, inexpensive reagents and equipment, safety, and reproducibility.\cite{synthesis-lead-flux, synthesis-lattice-parameters, synthesis-V2O5}
Synchrotron powder XRD for DyVO$_4$ was carried out at the Advanced Photon Source beamline 11-BM, with calibrated X-ray wavelength $\lambda=0.458057$~{\AA}.
Additionally, powder XRD for DyVO$_4$ and solid solutions was also performed using a laboratory Bruker D8 ADVANCE diffractometer with a Mo X-ray tube.
Rietveld refinements for all diffraction data were performed with GSAS-II.\cite{GSASII}
Heat capacity measurements were performed using a Quantum Design Physical Property Measurement System with pressed pellets with details in Supplementary Materials.\cite{supplement}
Different cooling procedures were tested for pure DyVO$_4$, including 10~$^\circ$C/h, 150~$^\circ$C/h and air quenching, but both synchrotron powder XRD and heat capacity measurements show no difference among them, with details in Supplementary Materials.\cite{supplement}

\begin{figure}
\centering
\includegraphics[width=\columnwidth]{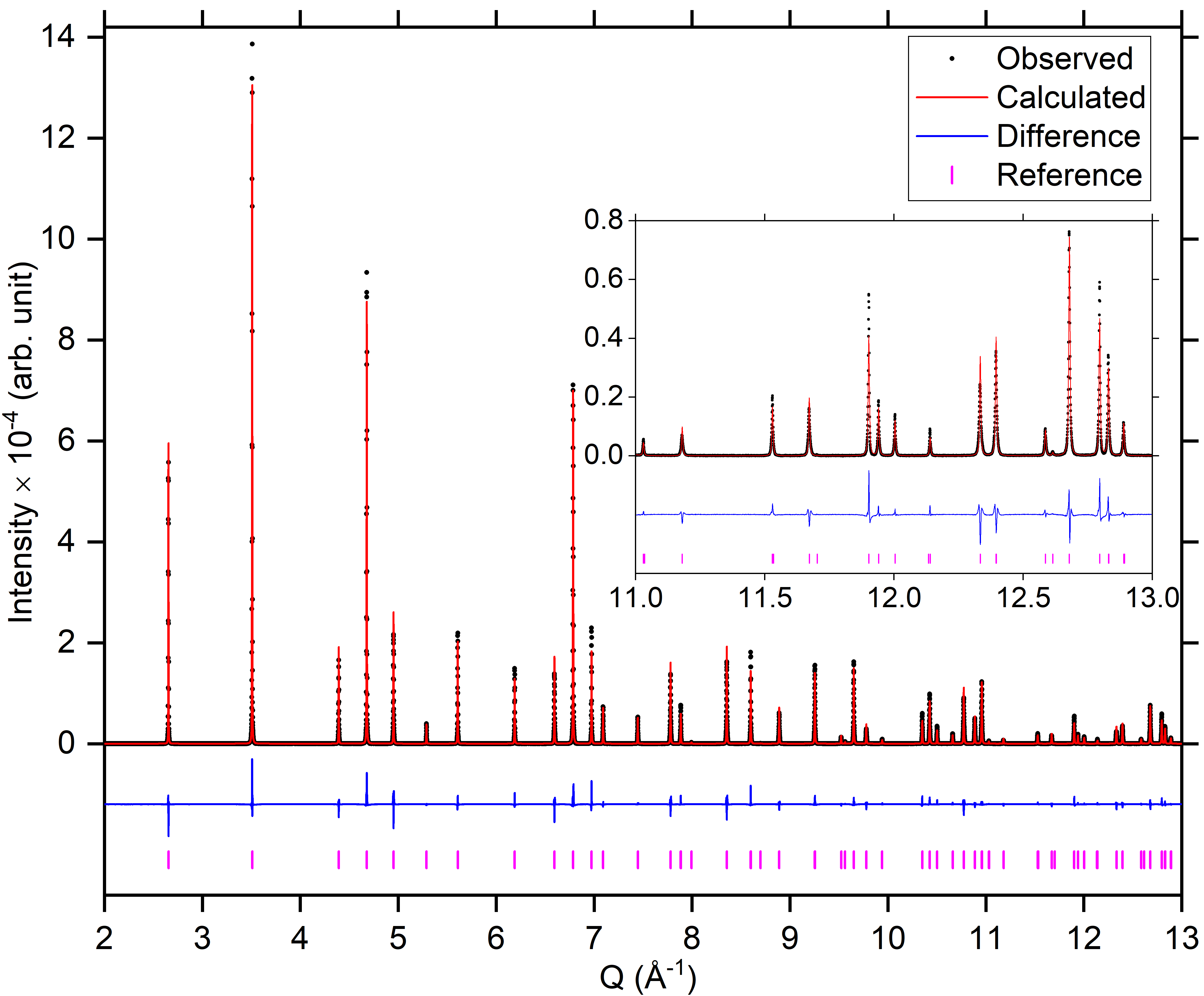}
\caption{Rietveld refined synchrotron powder XRD pattern of pure DyVO$_4$. The inset is a zoomed-in view of the 11~\AA$^{-1}$ to 13~\AA$^{-1}$ region.}
\label{fig: 11bm}
\end{figure}

\begin{figure*}
    \centering
    \includegraphics[width=\textwidth]{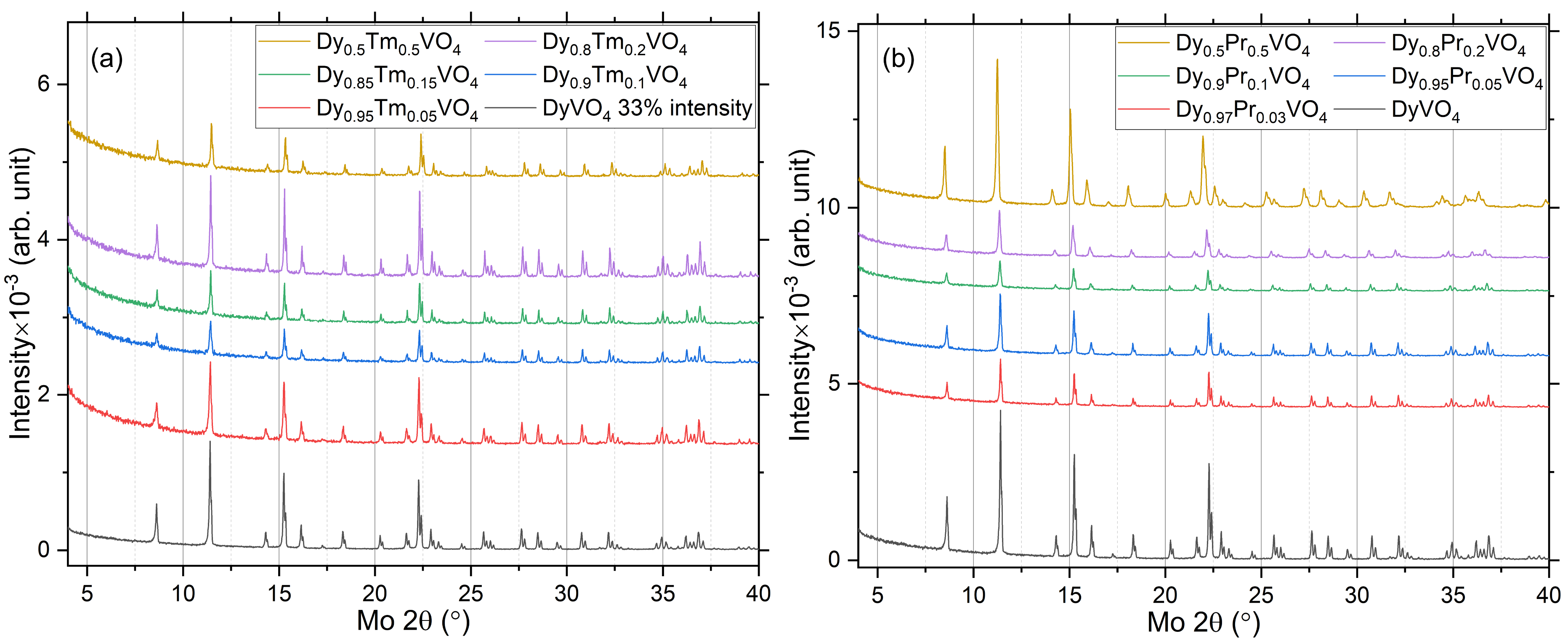}
    \caption{Stacked room-temperature powder XRD patterns for (a) Dy$_{1-x}$Tm$_x$VO$_4$ and (b) Dy$_{1-y}$Pr$_y$VO$_4$, together with DyVO$_4$, where the intensity of DyVO$_4$ pattern is rescaled to 33\% of its original value in (a).}
    \label{fig: stacked XRD}
\end{figure*}

Rietveld refinements for room-temperature synchrotron powder XRD shown in Figure \ref{fig: 11bm} demonstrate that DyVO$_4$ is phase pure with sharp peaks. The refined lattice parameters are: $a=7.139208(21)$~{\AA}, and $c=6.299997(18)$~{\AA} for tetragonal DyVO$_4$, with Rw = 15.079\%, very close to previously reported values.\cite{DyVO4-lattice-parameters-splitting}

For solid solutions, stacked XRD patterns are shown in Figure \ref{fig: stacked XRD}, where the pattern of DyVO$_4$ is included as a reference.
Compared to pure DyVO$_4$, no extra peaks or peak splittings can be seen for solid solutions, which demonstrates the successful formation of tetragonal solid solutions, instead of a stoichiometric mixture between DyVO$_4$ and TmVO$_4$ or PrVO$_4$.
In Figure \ref{fig: stacked XRD}(b), lattice expansion with Pr$^{3+}$ is most apparent from peaks shifting to lower 2$\theta$, especially at higher angles. 
This expansion and shifting is also apparent in Figure \ref{fig: stacked XRD}(a) for Tm$^{3+}$ substitution, though to a lesser extent due to its smaller size mismatch with Dy$^{3+}$ compared to Pr$^{3+}$.

\begin{figure*}
\centering
\includegraphics[width=\textwidth]{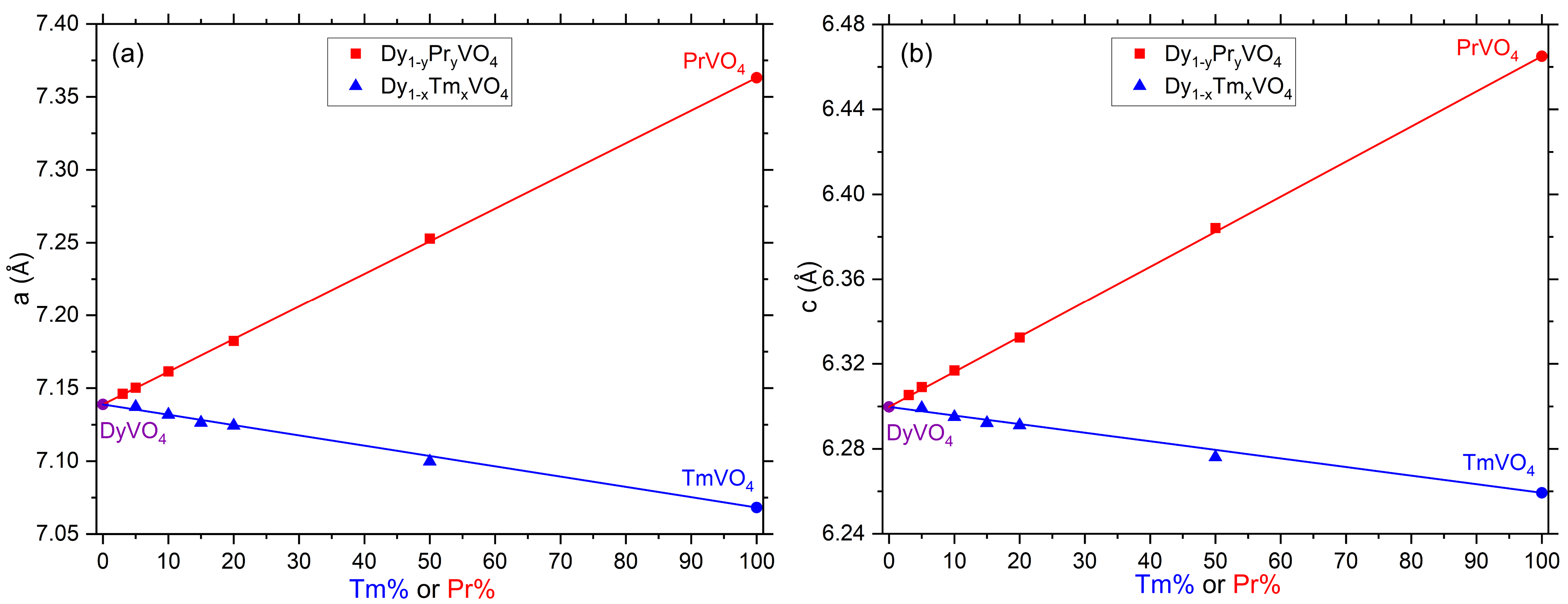}
\caption{Rietveld refined lattice parameters $a$ in (a) and $c$ in (b). The Dy$_{1-x}$Tm$_x$VO$_4$ system is denoted by blue triangles and the Dy$_{1-y}$Pr$_y$VO$_4$ system by red squares. The round dots represent pure DyVO$_4$, TmVO$_4$, and PrVO$_4$ compounds, and the straight lines show Vegard's law.}
\label{fig: lattice parameters}
\end{figure*}

Rietveld refinements were done for all the patterns shown in Figure \ref{fig: stacked XRD} to obtain the lattice parameters $a$ and $c$ for tetragonal solid solutions, with the results shown in Figure \ref{fig: lattice parameters}.
The lattice parameters of the end members PrVO$_4$ and TmVO$_4$ are obtained from Chakoumakos, et al.\cite{whole-group-new-TmVO4-PrVO4} but the lattice parameters of DyVO$_4$ are Rietveld refined.
The points representing DyVO$_4$ and either TmVO$_4$ or PrVO$_4$ are connected by straight lines, showing that both lattice parameters obey Vegard's law of mixing closely,\cite{Vegard's-law} which  further demonstrates the continuous substitution and preservation of the tetragonal zircon structure.

With pure DyVO$_4$ and two solid solutions systems successfully made, heat capacity measurements were carried out using a piece cut from a pressed pellet for each material, to determine the JT transition temperatures for each material.
The heat capacity measurements for pure DyVO$_4$ over the broad range from 2.5~K and 200~K is shown in Figure S4 in Supplemental Materials,\cite{supplement} showing the AFM transition at around 3~K and JT transition at 13.6~K.

\begin{figure*}
    \centering
    \includegraphics[width=\textwidth]{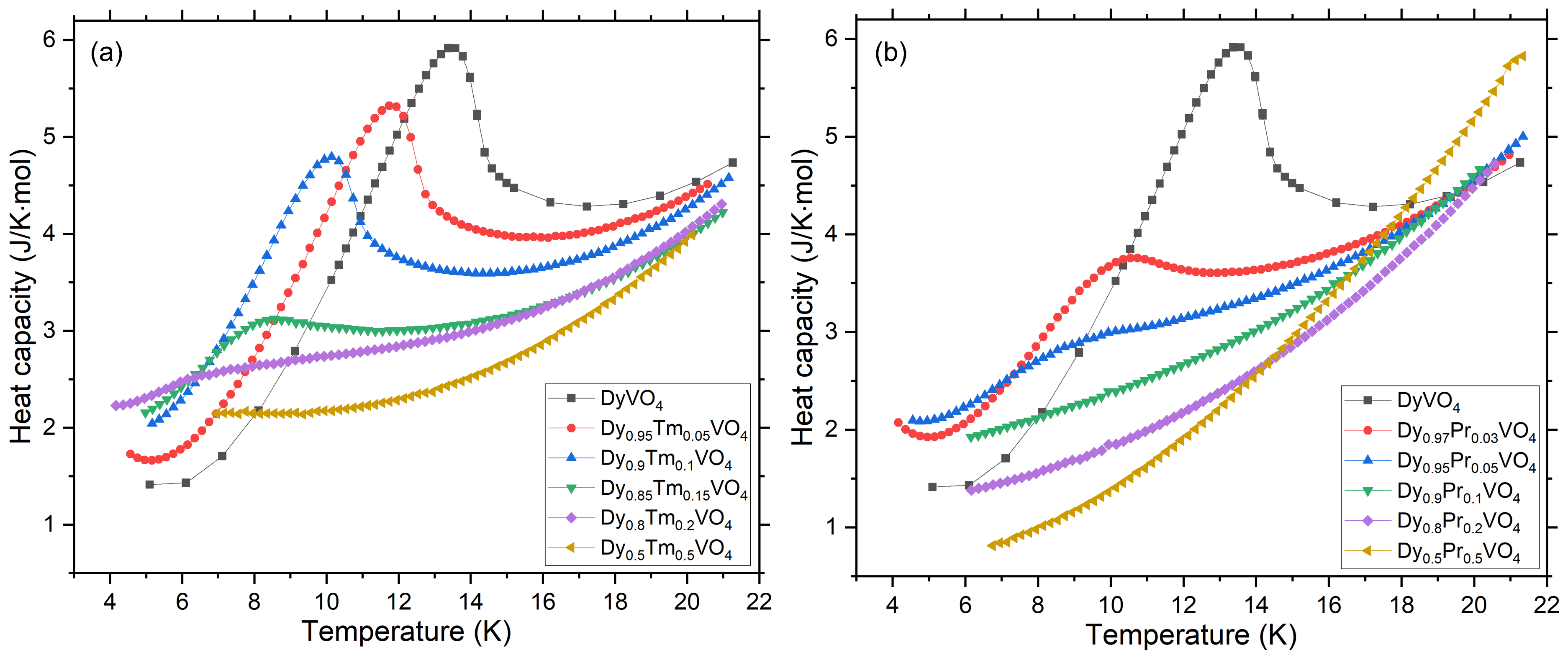}
    \caption{Heat capacity vs. temperature of (a) Dy$_{1-x}$Tm$_x$VO$_4$ and (b) Dy$_{1-y}$Pr$_y$VO$_4$, with DyVO$_4$ included.}
    \label{fig: heat capacity}
\end{figure*}

For solid solutions, the heat capacity measurements focused on the temperature range containing the JT transition peak.
The heat capacity curves of Dy$_{1-x}$Tm$_x$VO$_4$ and Dy$_{1-y}$Pr$_y$VO$_4$ systems are shown in Figure \ref{fig: heat capacity}.
Here we choose Tm$^{3+}$ and Pr$^{3+}$ as replacements of Dy$^{3+}$ because TmVO$_4$ also undergoes a cooperative JT transition at $T_\mathrm{JT}=2.1$~K,\cite{TmVO4-1, TmVO4-2} while PrVO$_4$ does not undergo cooperative JT transition. As a result, should we expect the transition temperature to smoothly vary for Dy$_{1-x}$Tm$_x$VO$_4$, but disappear eventually for Dy$_{1-y}$Pr$_y$VO$_4$?

The heat capacity curve of DyVO$_4$ is also included in each figure as a reference, using the same sample as shown in Figure \ref{fig: 11bm}.
DyVO$_4$ has a JT transition peaked at around 13.6~K, as previously reported.\cite{DyVO4-heat-capacity-Ising, DyVO4-heat-capacity-2, DyVO4-heat-capacity-3}
Published reports of heat capacity in DyVO$_4$ can exhibit sharper peaks if they are conducted on single crystals, as opposed to pressed pellets as we examine here. 
For solid solutions, the JT transition temperature is suppressed lower, along with the weakening peak height, and eventually, the peak ceases to exist when the composition of Tm$^{3+}$ or Pr$^{3+}$ is high enough.
For the Dy$_{1-x}$Tm$_x$VO$_4$ system, approximately 20\% of Tm$^{3+}$ suppresses the JT transition, while for the Dy$_{1-y}$Pr$_y$VO$_4$ system, even 5\% of Pr$^{3+}$ causes the JT transition peak to disappear.

Mean-field theory has traditionally been used to explain cooperative JT transitions,\cite{Raman-review} even for solid solutions.\cite{solid-solution}
Each JT-active ion is considered to be in a mean field, which is the average of the interaction from all neighbors.
The four lowest nearly-degenerate levels in DyVO$_4$ can be considered as a quadrupole,\cite{solid-solution} and JT transition can be explained by the alignment of quadrupoles using the Ising model with the introduction of pseudo-spin $\sigma^z(i)=\pm1$ for each JT-active site $i$.\cite{DyVO4-heat-capacity-Ising, Gehring-long}
The high-temperature crystal field splitting of the two Kramers doublets of Dy$^{3+}$ with $2\epsilon=9~\mathrm{cm^{-1}}$ acts like a transverse field, so that the Hamiltonian for DyVO$_4$ with no external magnetic field is
\begin{equation}
    H=\frac{1}{2}\sum_{i, j} J(i, j)\sigma^z(i) \sigma^z(j) + \sum_{i} \epsilon \sigma^x(i).
\end{equation}
The first term sums the exchange energies $J(i,j)$ in an Ising model, and the second term is the crystal field energy.
For each JT-active site $i$, the effect from all other interacting sites can be approximated by a mean field 
\begin{equation}
    \lambda\langle\sigma^z\rangle=\frac{1}{2}\sum_{j} J(i, j) \sigma^x(j),
\end{equation}
where $\langle\sigma^z\rangle$ is the order parameter, and the Hamiltonian will have the energy levels of
\begin{equation}
    \pm W=\pm (\lambda^2{\langle\sigma^z\rangle}^2+\epsilon^2)^{1/2}.
\end{equation}
The self-consistency equations for the energy and the order parameter $\langle\sigma^z\rangle$ in mean-field approximation would be
\begin{equation} \label{eq:self1}
    W=\lambda \tanh{\frac{W}{k_\mathrm{B}T}},
\end{equation}
and
\begin{equation}\label{eq:self2}
    \langle\sigma^z\rangle=\frac{{\lambda}{\langle\sigma^z\rangle}}{W} \tanh{\frac{W}{k_\mathrm{B}T}},
\end{equation}
where Equations (\ref{eq:self1}) and (\ref{eq:self2}) are equivalent.
The order parameter $\langle\sigma^z\rangle$ would be 1 at $T=0~\mathrm{K}$, meaning all the quadrupole moments are aligned. $\langle\sigma^z\rangle$ would be 0 at the JT transition temperature $T_\mathrm{JT}$, which is the point where quadrupole moments are still randomized, but the alignment is about to happen. Thus by letting $\langle\sigma^z\rangle=0$ Equation (\ref{eq:self2}), we have
\begin{equation}
    {\frac{\lambda}{\epsilon}} \tanh{\frac{\epsilon}{k_\mathrm{B}T_\mathrm{JT}}}=1.
    \label{important}
\end{equation}

For solid solutions, the mean-field strength $\lambda$ would decrease proportionally with decreasing concentration of Dy$^{3+}$.
A cooperative JT effect relies on interaction of JT-active Dy$^{3+}$ sites with each other, not with other rare earth ions.
As a result, replacing Dy$^{3+}$ with Tm$^{3+}$ or Pr$^{3+}$ will disturb the original Dy$^{3+}$ lattice, thus decreasing the mean-field strength $\lambda$.
Also, an internal strain energy will be induced as Pr$^{3+}$ (ionic radius 1.13~\AA) or Tm$^{3+}$ (ionic radius 1.02~\AA) ions substitutes Dy$^{3+}$ (ionic radius 1.052~\AA) ions, due to the mismatch of ion sizes.\cite{ionic-radii}
The strain energy can be considered as a transverse field as well, similar to the crystal splitting energy in the Hamiltonian, thus introducing another orthogonal term in the Hamiltonian. The Hamiltonian after mean-field approximation now becomes
\begin{equation}
    H=\sum_{i} \lambda\langle\sigma^z\rangle \sigma^z(i) + \sum_{i} \epsilon \sigma^x(i) + \sum_{i} \epsilon^\prime \sigma^y(i),
\end{equation}
where the final $\epsilon^\prime$ term denotes the internal strain energy due to lattice mismatch.
This will effectively increase $\epsilon$ in Equation (\ref{important}), by replacing $\epsilon$ with $\sqrt{\epsilon^2+{\epsilon^\prime}^2}$.

Let us focus on the Dy$_{1-x}$Tm$_x$VO$_4$ system. As $x$ increases, the mean-field would decrease to $(1-x)\lambda$ in Equation (\ref{important}), but $\epsilon$ increases, so the pre-factor in front of the hyperbolic tangent decreases. Since the hyperbolic tangent changes slowly with its argument, the transition temperature $T_\mathrm{JT}$ must decrease disproportionately to make the argument increase dramatically.
This decrease of $T_\mathrm{JT}$ is shown in heat capacity measurements, and $T_\mathrm{JT}$ indeed decreases disproportionately with $x$.
This shift is easily tunable but is obviously more rapid than a naive model where the system linearly varies between TmVO$_4$ and DyVO$_4$.
Eventually, as $x$ becomes large enough, the transition temperature is pushed down to zero, and JT transition will not happen.

From an electron energy point of view, as $x$ increases and $(1-x)\lambda$ decreases, the energy gain due to cooperative JT transition would decrease. However, the strain energy would increase with $x$. 
When $x$ is large enough such that there is no net energy gain from JT transition, the JT transition is totally suppressed.
If the above argument is applied to Figure \ref{fig: DyVO4-levels}, then the crystal field splitting above 13.6~K would increase as $x$ increases.
However, the energy gain due to the JT transition below 13.6~K, which is originally 18~cm$^{-1}$ for the 6 relevant electrons, would decrease. Thus, the decreasing net energy gain from JT transition will result in a decreasing latent heat, shown in heat capacity measurements. Additionally, there must be a point where the splittings above and below $T_\mathrm{JT}$ become equal and the JT transition will cease to happen.

The same argument can be used to explain the JT transition temperature and peak height trend in Dy$_{1-y}$Pr$_y$VO$_4$.
Comparing these two systems, for the same level of substitution, such as $x=y=5\%$, the Dy$_{1-y}$Pr$_y$VO$_4$ system has a lower $T_\mathrm{JT}$ and a lower peak height.
The strain energy is larger in Dy$_{1-y}$Pr$_y$VO$_4$ system due to the larger mismatch of ionic radii between Dy$^{3+}$ and Pr$^{3+}$.\cite{ionic-radii}
Thus for the same $x$ and $y$, the mean-field strength $\lambda$ would be the same, but the $\epsilon^\prime$ for Dy$_{1-y}$Pr$_y$VO$_4$ system would be larger hence its $T_\mathrm{JT}$ and peak height are both lower.

Mean-field theory captures the trends in these systems but does not precisely predict the JT transition temperature for each material since the short range interaction of Dy$^{3+}$ is very strong.\cite{mean-field-short-range, DyVO4-heat-capacity-Ising} 
In the above analysis, addition of Tm$^{3+}$ or Pr$^{3+}$ only reduces the number of JT-active Dy$^{3+}$ sites while providing strain--they do not alter the electronic configuration of Dy$^{3+}$ sites (Figure \ref{fig: DyVO4-levels}).
If the proportion of Tm$^{3+}$ ions becomes large enough so that they can interact with each other cooperatively, the JT transition of TmVO$_4$ (with a minority of Dy$^{3+}$ ions) is expected, which results in a ``valley'' with no JT transition for intermediate substitutions, but this will not happen for PrVO$_4$.

In conclusion, we demonstrate the successful synthesis of pure DyVO$_4$ and solid solutions Dy$_{1-x}$Tm$_x$VO$_4$ ($x$ = 0.05, 0.1, 0.15, 0.2, and 0.5) and Dy$_{1-y}$Pr$_y$VO$_4$ ($y$ = 0.03, 0.05, 0.1, 0.2, and 0.5) by XRD and Rietveld refinements.
Both solid solution systems have decreasing cooperative JT transition temperatures as the substitution level $x$ or $y$ increases, shown in heat capacity measurements, with decreasing latent heats. 
The trend of $T_\mathrm{JT}$ for the solid solution systems is explained by mean-field approximation.
This model explains that the cooperative JT effect is suppressed for high substitution levels, even when the end members both exhibit JT transitions, such as DyVO$_4$ and TmVO$_4$.
In diffraction experiments, the JT transition is evident and peak splitting is expected.\cite{DyVO4-lattice-parameters-splitting}
Due to the low and well-defined transition temperatures, these non-hazardous rare earth vanadates and their solid solutions with cooperative JT transition can be used as temperature calibrants for diffraction experiments at cryogenic temperatures, as an alternative to diodes or thermocouples that require sample contact. 

\begin{acknowledgments}
This work was supported by the US Department of Energy, Basic Energy Sciences (grant No.\ DE-SC0013897) for Early Career Research.
\end{acknowledgments}

\section*{Data Availability Statement}

The data that support the findings of this study are available from the corresponding author upon reasonable request.

\bibliography{REVO4}

\end{document}



\begin{center}
\Large 
\textbf{Tuning of co-operative Jahn-Teller transition temperature in rare earth vanadate solid solutions}\\
\vspace{10mm}
Supplementary Material\\
\vspace{10mm}
\normalsize
Kejian Qu, Daniel P. Shoemaker\\
\vspace{10mm}
\Large 
\textbf{Contents}
\end{center}

\begin{enumerate}
  \item Experimental methods: detailed synthesis and measurement procedures, together with the comparison of DyVO$_4$ samples with different cooling rates.
  \item Figure S1: comparison of synchrotron XRD for DyVO$_4$ with different cooling rates.
  \item Figure S2: zoomed-in view of synchrotron XRD comparison.
  \item Figure S3: comparison of heat capacity curves for DyVO$_4$ with different cooling rates.
  \item Figure S4: heat capacity curve of DyVO$_4$ between 2.5~K and 200~K.
\end{enumerate}

\pagebreak
\begin{center}
    \textbf{Experimental Methods}
\end{center}

\textit{(1) Synthesis:} pure DyVO$_4$ together with solid solutions Dy$_{1-x}$Tm$_x$VO$_4$ (x = 0.05, 0.1, 0.15, 0.2, and 0.5) and Dy$_{1-y}$Pr$_y$VO$_4$ (y = 0.03, 0.05, 0.1, 0.2, and 0.5) were all synthesized by solution precipitation.
Stoichiometric Dy(NO$_3$)$_3${$\cdot$}5H$_2$O (99.9\%), Pr(NO$_3$)$_3${$\cdot$}5H$_2$O (99.9\%) or Tm(NO$_3$)$_3${$\cdot$}5H$_2$O (99.9\%), together with V$_2$O$_5$ (99.6+\%) were mixed and nitric acid (15.8 normality) was added to the mixture.
Extra nitric acid does not affect the synthesis.
The liquid samples were heated to 100~$^\circ$C on a hot plate while stirring for 1 hour to allow fully mixing, and then they were transferred to drying dishes.
Some undissolved V$_2$O$_5$ residue was rinsed with nitric acid to make sure the stoichiometry is preserved.
The drying dishes was dried on the hot plate at 100~$^\circ$C until no liquid was left, for around 12 hours.
The precipitate samples were scraped off the drying dish and fully ground in mortar and the powder was loaded into alumina crucibles.
These crucibles were heated to 900~$^\circ$C in 9 hours in air, and stayed at 900~$^\circ$C for 36 hours.
The final product of each sample was ground into fine powder (around 1~$\mu$m) and collected, the total mass of each sample was around 1~g to 2~g.

\textit{(2) Cooling procedures:} different cooling rates were tested for pure DyVO$_4$, including 10~$^\circ$C/h, 150~$^\circ$C/h(only for temperature above 150~$^\circ$C, cooling naturally with furnace when temperature dropped below 150~$^\circ$C), and air quenching.
Figure S1, S2, S3 compare DyVO$_4$ synthesized by different cooling rates.
For all the solid solutions, the samples were cooled down naturally with the furnace because the sample quality did not change with varying cooling rates.

\textit{(3) Diffraction measurements:} for DyVO$_4$ with different cooling rates, room-temperature synchrotron powder XRD was performed at beamline 11-BM of the Advanced Photon Source in Argonne National Laboratory through the sample mail-in service.
The powder samples were sealed in kapton capillaries and the calibrated wavelength is 0.458057~{\AA}.
Additionally, powder XRD for pure DyVO$_4$ and all the solid solutions was performed using an in-lab Bruker D8 Advance diffractometer with a Molybdenum X-ray source (K$\alpha1=0.71$~{\AA}) at room temperature with samples sealed in kapton capillaries.
Rietveld refinements for all diffraction data were done with GSAS-II.

\textit{(4) Heat capacity measurements:} for each material, a dense pellet was pressed from powder and a small piece around 7~mg to 8~mg was cut from the pressed pellet for heat capacity measurements.
The measurements were carried out with Quantum Design Physical Property Measurement System (PPMS) in Materials Research Laboratory in the University of Illinois, from around 20~K to the lowest possible temperature the PPMS was able to achieve, usually around 5~K, and this temperature range covers all the transition peaks.
For DyVO$_4$ with different cooling rates, heat capacity measurements were done between 11~K and 16~K.

\begin{figure}[h]
    \centering
    \includegraphics[width=0.9\columnwidth]{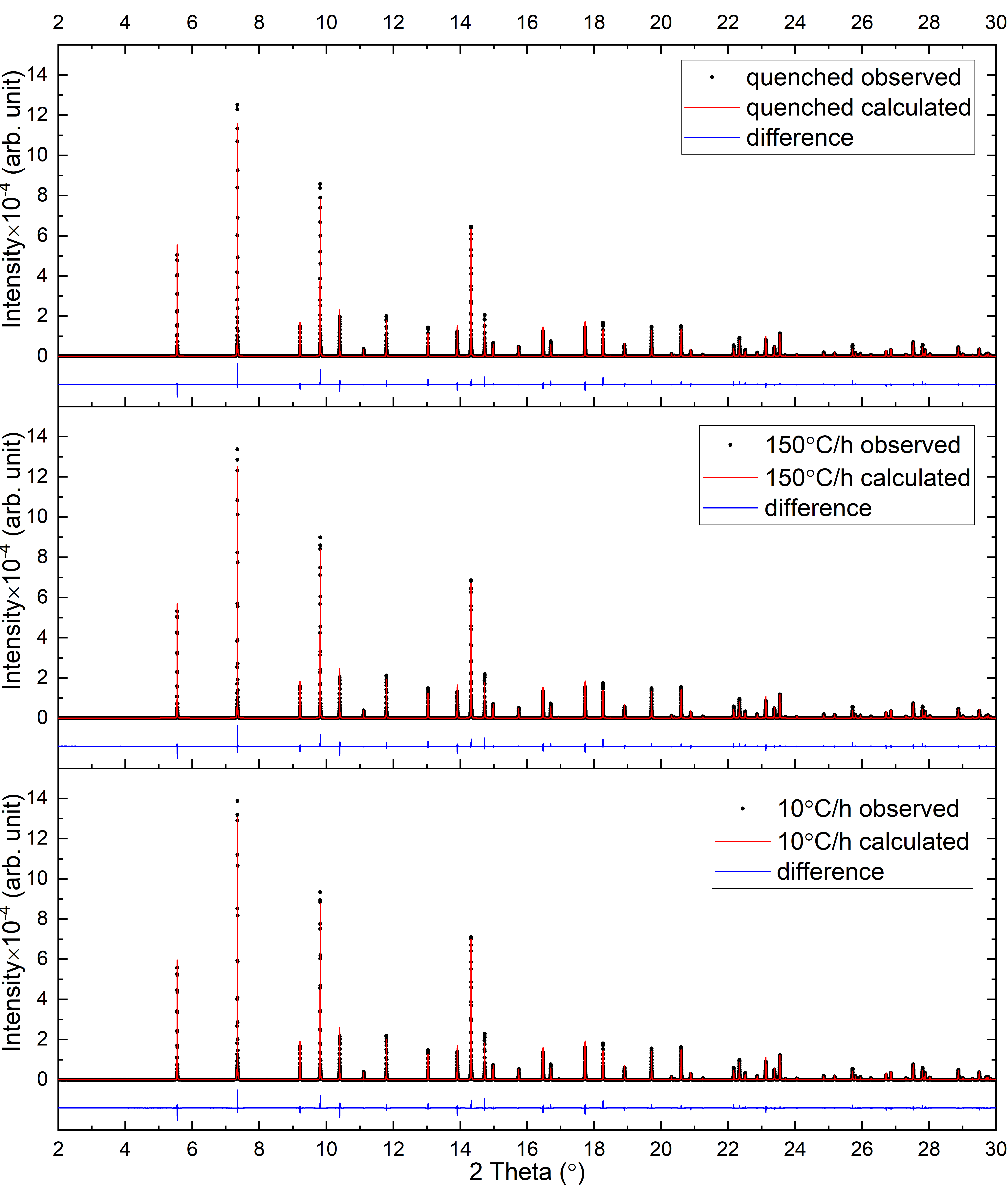}
    \caption{Comparison of room temperature 11-BM XRD patterns of DyVO$_4$, with calibrated wavelength $\lambda=0.458057$~\AA. From bottom to top, the samples are slow-cooled (10~$^\circ$C/h), fast-cooled (150~$^\circ$C/h), and quenched, respectively. And the bottom panel is the same as Figure 2 in the main text. No obvious peak broadening can be seen. The Rietveld refined microstrains are 626.3 ($R$w=15.079\%), 629.2 ($R$w=14.551\%), and 625.7 ($R$w=13.941\%), for slow-cooled, fast-cooled and quenched samples, respectively, and the microstrain is essentially the same across the samples. To visualize the peak sharpness better, the pattern between 10$^\circ$ and 14$^\circ$ is zoomed-in and shown in Figure 3.}
    \label{fig: 11bm comparison}
\end{figure}

\begin{figure}
    \centering
    \includegraphics[width=0.9\columnwidth]{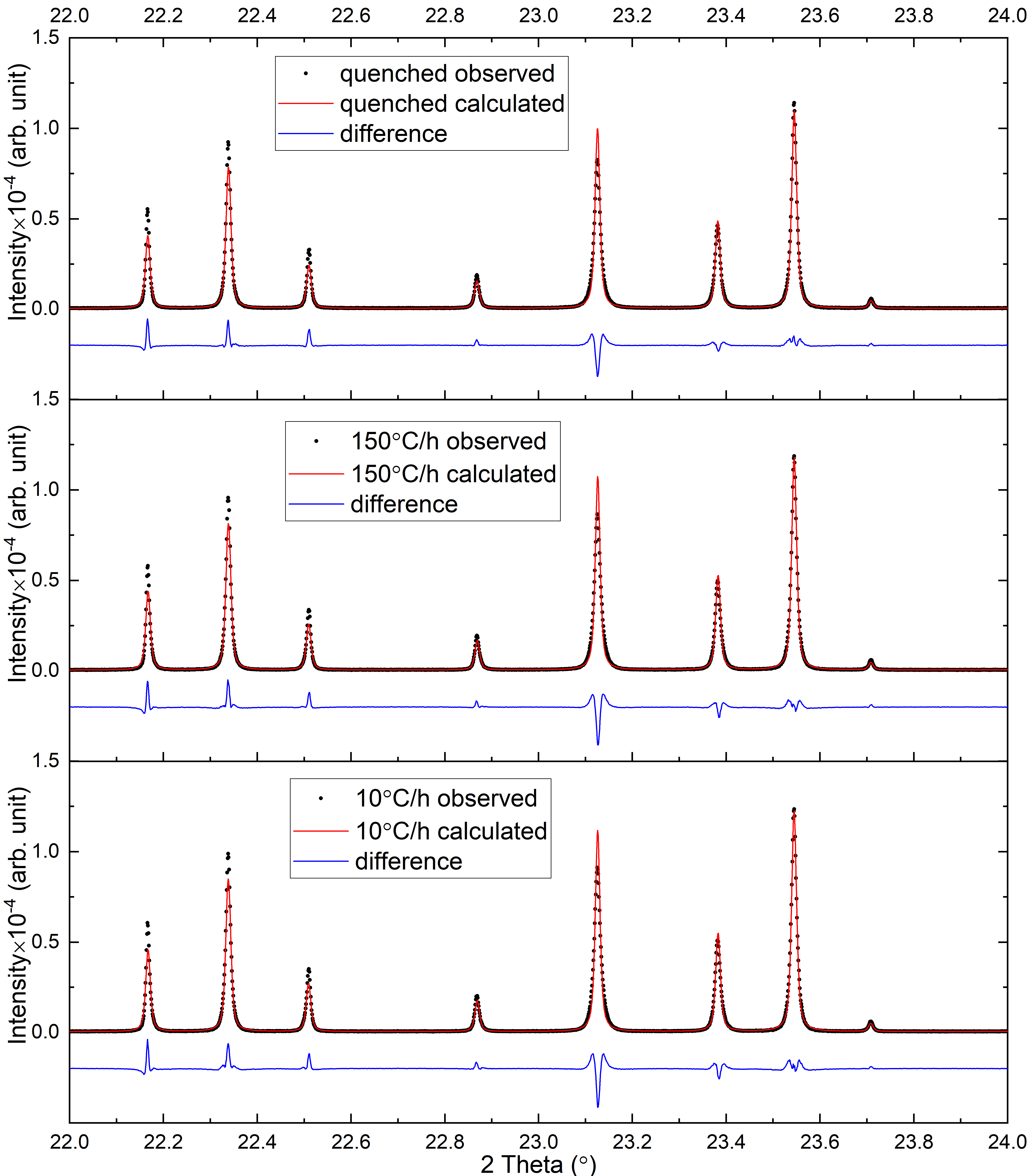}
    \caption{Comparison of zoomed-in room temperature 11-BM XRD patterns of DyVO$_4$, with calibrated wavelength $\lambda=0.458057$~\AA, from 2$\theta=10^\circ$ to 2$\theta=14^\circ$. From bottom to top, the samples are slow-cooled (10~$^\circ$C/h), fast-cooled (150~$^\circ$C/h), and quenched, respectively. No obvious peak broadening can be seen for these peaks.}
    \label{fig: 11 bm comparison zoom in}
\end{figure}

\begin{figure}[h]
    \centering
    \includegraphics[width=0.9\columnwidth]{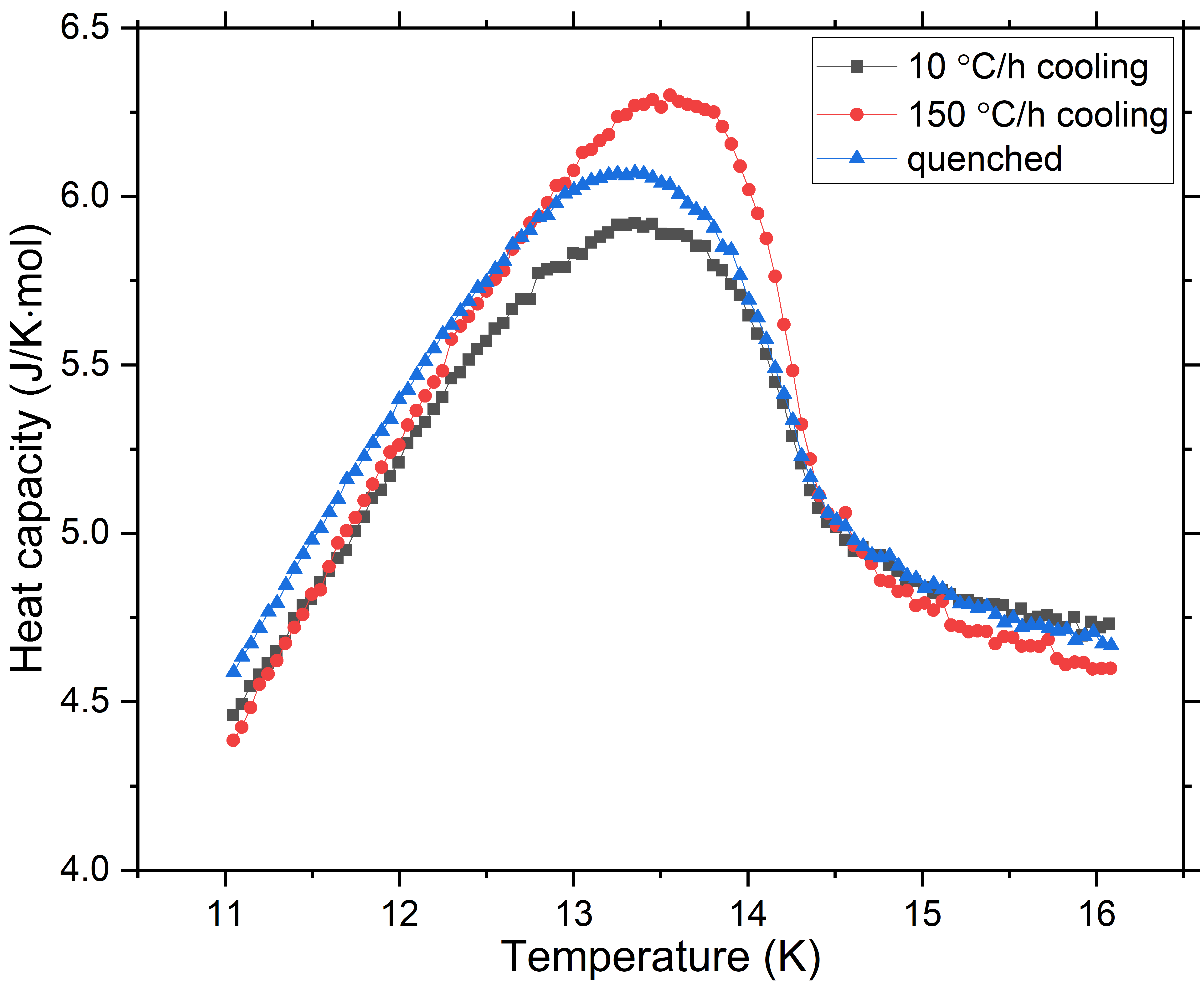}
    \caption{Comparison of heat capacity measurements for DyVO$_4$ samples with different cooling rates near the JT transition temperature. No systematic change of the shape of the curves can be seen across the samples. The slow-cooled (10~$^\circ$C/h) DyVO$_4$ sample is used for 11-BM diffraction measurements in the main text.}
    \label{fig: heat capacity comparison}
\end{figure}

\begin{figure}[h]
    \centering
    \includegraphics[width=0.9\columnwidth]{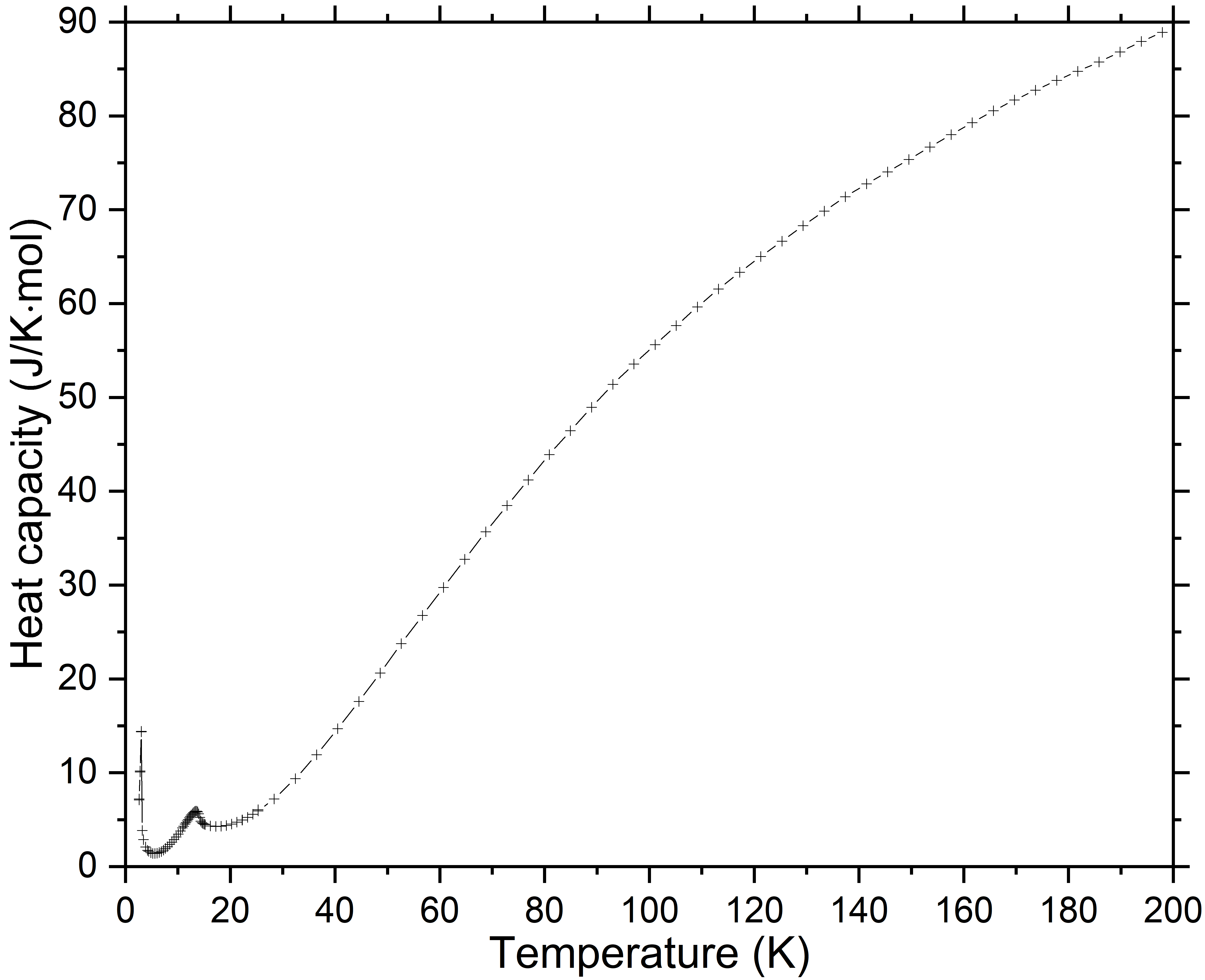}
    \caption{Heat capacity of DyVO$_4$ between 2.5~K and 300~K. The peak at around 14~K is the JT transition peak, while the peak at 3~K is the antiferromagnetic transition peak.}
    \label{fig: heat capacity comparison}
\end{figure}